\documentclass[twocolumn,showpacs,aps,prl,superscriptaddress]{revtex4}

\usepackage{graphicx}
\usepackage{dcolumn}
\usepackage{bm}

\begin{document}

\title{
Role of the spin-orbit splitting and the dynamical fluctuations
in the Si(557)-Au surface.
}

\author{Daniel S\'anchez-Portal}
\email{sqbsapod@sc.ehu.es}
\author{Sampsa Riikonen}
 \affiliation{ Unidad de F\'{\i}sica de Materiales CSIC-UPV/EHU,
Donostia International Physics Center (DIPC), and
Departamento de F{\'{\i}}sica de Materiales,
Universidad del Pa\'{\i}s Vasco,
Apdo. 1072, 20080 Donostia, Spain }
\author{Richard M. Martin}
\affiliation{
Deparment of Physics and Materials Research Laboratory,
University of Illinois, Urbana, IL 61801, USA}


\begin{abstract}

\centerline{\bf Accepted for publication in {\it Phys. Rev. Lett.}
, August 2004}

Our {\it ab initio} calculations show that spin-orbit coupling is
crucial to understand the electronic structure of the Si(557)-Au
surface. The spin-orbit splitting produces the two one-dimensional
bands observed in photoemission, which were previously attributed
to spin-charge separation in a Luttinger liquid. This spin
splitting might have relevance for future device applications. We
also show that the apparent Peierls-like transition observed in
this surface by scanning tunneling microscopy is a result of the
dynamical fluctuations of  the step-edge structure, which are
quenched as the temperature is decreased.
\end{abstract}


\maketitle
Atomic-scale ``wires'' created on semiconductor surfaces
have the potential to be the basis for nanoscale devices and
to demonstrate the novel physics of one dimension.
A specially attractive route to create such structures
is provided by the
use of stepped
vicinal surfaces,
which can be used as templates
to deposit material
creating very regular arrays of parallel
metallic chains~\cite{Himpsel04}.
The Si(557)-Au reconstruction is one of these systems.
The terraces of this surface are perpendicular to the (111) direction
and have a
width of $\sim$19~\AA. Each terrace contains a
monatomic chain of gold atoms running parallel to the step-edge.
For this reason the Si(557)-Au surface has been
proposed as an experimental realization
of a one-dimensional
metal and has attracted much attention in recent
years~\cite{Segovia99,Losio01,Robinson02,Ahn03,Crain03,PRB02,SurfSci03}.
However, in spite of these efforts, the electronic structure of this system
is not yet completely understood.

The first angle-resolved photoemission (ARP)
study by Segovia {\it et al.}\cite{Segovia99}
found a spectrum dominated by a one-dimensional metallic band.
This band was shown to
split in two peaks
near the Fermi level (E$_F$), and this was interpreted as signature
of separated charge and spin low-energy excitations as predicted
by Tomonaga-Luttinger theory of the one-dimensional
electron gas~\cite{Tomonaga,Luttinger}.
However, later photoemission data seemed
to discard this interpretation. According to
Losio {\it et al.}~\cite{Losio01}
the observed splitting
would correspond to
two distinct proximal bands which cross E$_F$
at neighboring, although different, positions of the surface
Brillouin zone.
However, the origin of these bands was unclear.
Finally,
Ahn {\it et al.}~\cite{Ahn03} have recently suggested that only
one of the bands is truly metallic and
suffers a metal-insulator transition upon cooling. This observation
was correlated with the temperature dependence
of the scanning tunneling microscopy (STM) images:
the step-edge
undergoes a
periodicity doubling consistent with a Peierls-like instability.
Thus these authors concluded that at least
one of the two proximal bands should be associated with
the atoms forming
the step-edge.

In contrast with the
electronic structure,
the geometry of the Si(557)-Au reconstruction seems to be quite
well established.
A detailed model was recently proposed by
Robinson {\it et al.}~\cite{Robinson02} on the basis of
X-ray
diffraction data, which
has been corroborated
by first-principles density functional
(DFT) calculations~\cite{SurfSci03}.
Unfortunately,
the calculated band structure only presents {\it one}
one-dimensional metallic
band exhibiting a considerable dispersion and a width consistent
with the experimental observations.
Thus, the observed two-band photoemission spectrum remains
unexplained from a theoretical point of view.

In this Letter we present the results of DFT calculations that
satisfactorily explain the observed ARP spectra and the
temperature dependence of the
STM images of the Si(557)-Au surface.
The two proximal
bands appear as a consequence of the spin-orbit (SO)
splitting of the most
dispersive surface state.
The large effect of SO splitting on gold-derived surface states
has been demonstrated
previously by ARP experiments and calculations on
Au(111)~\cite{Lashell96,Petersen00}.
The origin and magnitude
of the SO splitting is similar in the present case.
The inversion symmetry is always broken at the surface, thus
the spin splitting of bands with no spatial
degeneracy becomes possible.
This suggests that
atomic wires formed by heavy atoms deposited
on semiconducting surfaces
could be used in the fabrication of
spin transistors~\cite{Datta90} and
spin-filter devices~\cite{Streda03}. Interesting phenomena can
also appear associated
with the competition between electron-electron interactions
and SO coupling in one-dimension~\cite{Hausler01,Governale02}.

Our results also show that the temperature dependence
of the STM images stems from the buckling of the step-edge,
whose atoms alternate between up and down positions.
A given "up-down" configuration
and the reverse one are separated by a
small energy barrier. At room temperature the step-edge
fluctuates between both structures, and the STM
only reflects the average surface electronic and atomic structure.
Upon cooling
the step-edge buckling can be revealed using STM, producing
the apparent doubling of the periodicity.

\begin{figure}
\includegraphics[keepaspectratio,width=7cm]{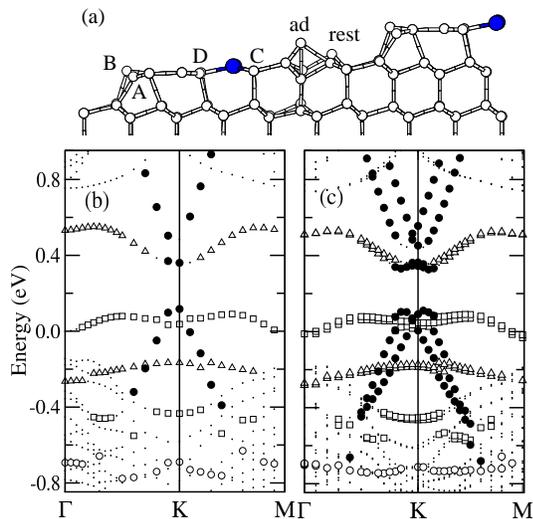}
\caption{\label{fig:fig1} (a) Calculated equilibrium structure
of the Si(557)-Au reconstruction.
The corresponding
electronic band structure is shown for a
calculation
not including (b), and including (c) the spin-orbit interaction.
Energies are referred to the Fermi level.
Surface states have been marked with different symbols according to
their main atomic character (see text).
}
\end{figure}

Our slabs contain four silicon bilayers, the bottom
surface saturated with hydrogen, with a total of 115 atoms.
Due to this large supercell we use the SIESTA
code~\cite{SIESTA1,SIESTA2} to study the energy
landscape and the structural properties of the Si(557)-Au surface.
The details of the calculations are similar to those
presented in Refs. \onlinecite{SurfSci03}
and \onlinecite{PRB02} for the same system.
Since the present version of
SIESTA does not allow to include the SO interaction,
we have used the VASP code~\cite{VASP,VASP2}
to study its effect
on the electronic structure of our relaxed
structures. The VASP calculations utilized projected-augmented-wave
potentials
with a plane-wave
cutoff of 250~eV, which proved to be sufficient to obtain
converged electronic band structures.

Fig.~\ref{fig:fig1}(a) shows the relaxed structure of the
Si(557)-Au reconstruction. This structure is almost identical to
the experimentally proposed model~\cite{Robinson02,SurfSci03}. The
larger, filled circles stand for the gold atoms occupying silicon
substitutional positions on the middle of the terraces. The
corresponding band structure, along the direction parallel to the
steps, is shown in Fig.~\ref{fig:fig1}(b) and (c). Panel (b) shows
the results from a non spin-polarized VASP calculation using the
local density approximation (LDA) for the exchange-correlation
potential. This band structure is almost identical to that already
reported in Ref.~\onlinecite{SurfSci03} using the SIESTA code.
Several surface bands and resonances can be identified, all of
them with negligible dispersion in the direction perpendicular to
the steps. The different symbols reflect their main atomic
character, as obtained from the projection of the charge onto
non-overlapping spheres centered on each atom. The unoccupied band
marked with open triangles comes from the adatoms (labeled {\it
ad} in Fig.~\ref{fig:fig1}(a)), while the occupied one is related
to the restatoms (labeled {\it rest}). In principle, every atom in
the step-edge has a dangling-bond pointing perpendicularly to the
step, which would give rise to a very flat half-filled band. This
unstable situation leads to a buckling of the step-edge that
doubles the unit cell and forms two bands marked by open squares
in Fig.~\ref{fig:fig1}.
The band with larger weight in the "up" (B) atoms
is fully occupied,
while the
band associated with the "down" (A) atoms has a
small
occupation~\cite{note}.
Notice that the step-edge bands have a very small dispersion. Thus
they cannot explain the observed ARP spectra,
as was recently suggested
by Ahn {\it et al.}~\cite{Ahn03}.
Contrary to the initial
interpretation of the ARP data~\cite{Segovia99}, none
of the surface bands in the proximity of E$_F$ has
a clear Au 6{\it s} character. This is a direct consequence of the
larger electron affinity
of gold as compared to silicon: the
6{\it s} Au character appears several eV below E$_F$.
The sole surface bands exhibiting
an appreciable gold component in Fig.~\ref{fig:fig1} are
marked with circles.
These bands mainly come
from the 3{\it p} states of the
silicon atoms neighboring to the Au chains (atoms D and C).
However, they also show a
strong contribution from the 6{\it p} states of gold, so they are
better assigned to the Si-Au bonds. We find a flat
band (open circles) associated with the Au-Si$_D$ bonds,
and a dispersive (filled circles) corresponding
to the Au-Si$_C$ bonds.
This last band is the {\it
only one} that presents a dispersion and width
consistent with the photoemission data.

\begin{figure}
\includegraphics[keepaspectratio,width=7.5cm]{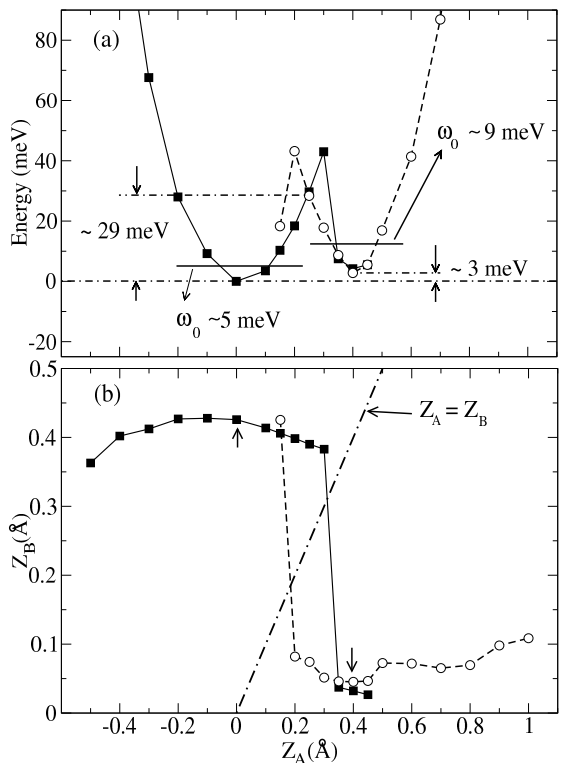}
\caption{\label{fig:fig2} Total energy (a)
and position of the B atom (b)
as a function of the height of atom A. Heights are referred
to that of atom A in the ground-state configuration. The arrows
in panel (b) mark the position of the two energy minima. The
energy landscape shows two wells separated by a small energy
barrier of $\sim$29~meV, and estimated
zero-point energies of $\omega_0\sim$ 5 and 9 meV.
}
\end{figure}

Fig.~\ref{fig:fig1}(c) depicts the same band structure once the effect
of the SO interaction has been included in a non-collinear spin
calculation. We still get a non spin-polarized
ground state.
The changes are negligible for most surface bands.
This is expected since most of them are localized
in regions far from the gold atoms or have
a very small dispersion, i.e. small group velocities.
However,
the dispersive Au-Si$_C$ band
develops a considerable SO splitting. This brings
the calculated band structure in reasonable agreement with
the experimental spectra.
The experiments~\cite{Losio01,Ahn03} 
show a $\sim$300~meV splitting near E$_F$.
This splitting exhibits a linear dependence as a function of
$k_\|$ with a
$\sim$1.2~eV \AA\ slope.
This can
be compared with our calculated $\sim$200~meV splitting at E$_F$
and $\sim$1.4~eV \AA\ slope.

Even though the two SO-split bands are a robust feature predicted
for this surface, the states at the Fermi energy are strongly
affected by the other bands shown in Fig.~\ref{fig:fig1}(b) and
(c). In our calculations both SO-split bands are metallic;
however, there is a band gap just above E$_F$. This gap relates to
the presence of a row of adatoms in the terrace, which induces an
alternation of the Si$_C$-Au-Si$_C$ bond angle between 101.8$^o$
and 109.6$^o$.
The presence of this gap drives
the surface very close
to becoming semiconducting. The metallicity of the system
is due to
the very small partial occupation of
the upper step-edge band that pins the position
of E$_F$.
A description of exchange
and correlation beyond LDA would quite likely cause a widening
of the step-edge gap, leading to a semiconducting
band structure at zero temperature.

\begin{figure}
\includegraphics[keepaspectratio,width=7cm]{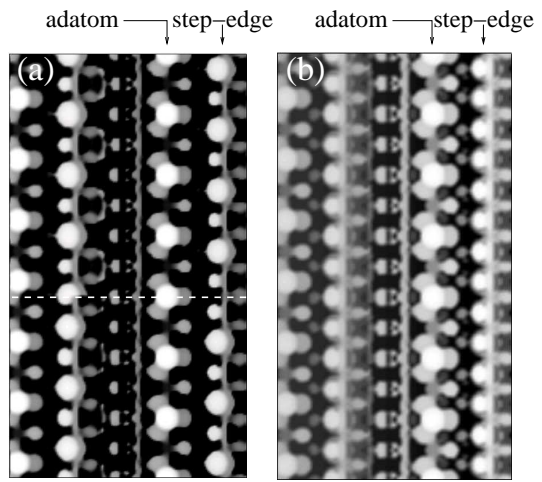}
\caption{\label{fig:fig3} Simulated STM images at low (a)
and high temperature (b) for a +0.7~V bias voltage (empty states).
The upper part of panel (a) corresponds to the ground state structure,
while the lower part is obtained with the reversed step-edge configuration.
Panel (b) combines both images.}
\end{figure}

The ground state structure and the electronic bands
shown in Fig.~\ref{fig:fig1} perfectly
explain
the low-temperature
STM images of the surface~\cite{Ahn03}.
The step-edge exhibits a buckling, with
A atoms lying $\sim$0.4~\AA\ below B atoms. Furthermore, a
high contrast between these two types of silicon atoms
is guaranteed.
Empty-state STM images will
preferentially show A atoms,
while filled-state images will reflect the location of B atoms.

The origin of the temperature dependence of the STM images is clarified in
Fig.~\ref{fig:fig2}, where the stability of the step-edge buckling
respect to thermal vibrations is analyzed using SIESTA.
Fig.~\ref{fig:fig2}
(a) shows the energy landscape
as a function of the position of atom
A, i.e. the lower atom of the
step-edge in the ground-state structure.
We progressively increase its height, Z$_A$
(see the filled symbols in Fig.~\ref{fig:fig2}).
At each step the whole structure is relaxed while keeping Z$_A$
as a constraint.
As expected the energy increases quadratically for
small displacements from the equilibrium position (Z$_A$=0).
The
height of the B atom (Z$_B$) remains almost
unchanged during this process as shown in Fig.~\ref{fig:fig2}(b) and,
although the magnitude of the gap decreases
as Z$_A$ approaches Z$_B$, we can still unambiguously identify
the fully and partially
occupied step-edge bands, respectively, with the
atoms B and A.
However, for larger displacements the situation is very different.
When Z$_A$$\geq$Z$_B$
the structure becomes unstable and collapses into a new configuration
with an inverted step-edge buckling.
Atom A is now situated $\sim$0.4~\AA\ above
atom B.
This sudden change is accompanied by a charge transfer
from atom B to atom A, so that the filled step-edge band can be always
assigned to the atom occupying the highest position.

The reversed buckling configuration is also
stable against small displacements
(see the open symbols in Fig.~\ref{fig:fig2}).
Its energy is only
$\sim$3~meV ($\sim$5~meV) larger
than that of the ground-state structure within the LDA
(generalized gradient
approximation, GGA). For large displacements we find a transition
back to the ground-state structure (open symbols in Fig.~\ref{fig:fig2}).
This means that the roles of atom A and B
can be interchanged without any
appreciable change of the global structure of the surface
and the total energy.
This is not surprising if we consider that, for the structure shown
in Fig.~\ref{fig:fig1}(a), the two step-edge sites only become
inequivalent when we consider their registry
with the adatom row in the same terrace, which is located at $\sim$12~\AA\
from the step-edge.
Thus the energy landscape as a function of the height of the
atoms in the step-edge is formed by two similar
wells.
The energy
barrier separating both structures turns out to be quite small,
$\sim$29~meV ($\sim$25~meV)
within LDA (GGA).

From the total energy curves and the corresponding displacements
of all the atoms in the supercell, the zero-point energies
are estimated to be
$\omega_0\sim$5~meV for the ground-state structure, and
$\sim$9~meV for the reversed buckling one. This, together with the
magnitude of the
calculated barrier, shows that
at low-temperatures both structures are indeed stable for long times.
The simulated images using
Tersoff-Hamann theory~\cite{TersoffHamann}
can be found in Fig.\ref{fig:fig3} (a), the upper part of the panel
corresponding
to the ground-state configuration and the lower part to the reversed
step-edge buckling.
The images are very similar. They are
dominated by two chains showing a double periodicity
along the step-edge direction in good agreement
with low-temperature images
reported for this system~\cite{Losio01,Ahn03}.
One of the chains is the row of adatoms.
The other corresponds to the step-edge, where only
every other atom
is visualized.

The situation changes at higher temperatures. The time
necessary to flip between the different
step-edge configurations gets shorter as the temperature increases.
Given the size
of the calculated energy barrier,
at room temperature the STM images will
only show a time average of the electronic
and atomic structure of the surface.
This has been modeled in Fig.~\ref{fig:fig3} (b)
by averaging the STM images of the two structures.
As a consequence, the double periodicity is lost
and all the atoms in the step-edge appear with similar intensities.
This explains the apparent periodicity doubling observed
by Ahn {\it et al.}~\cite{Ahn03}
upon cooling.
Therefore, the mechanism behind the
change of appearance
of the STM images of the Si(557)-Au surface
is similar to that proposed
to explain, for example, the
$3\times3$ $\rightarrow$
$\sqrt{3}\times\sqrt{3}$ transition
in Sn/Ge(111)~\cite{SnGe99}.

Our results also provide a plausible scenario to explain
the observation of a metal-insulator transition in this
surface~\cite{Ahn03}.
The Fermi energy - and thus the occupation and metallicity of the
SO-split bands - depends strongly upon the position of the upper
step-edge band. Small changes may drive the system from metallic
to insulator and conversely. Furthermore, as the temperature
increases the step-edge structure is expected to fluctuate at an
increasing rate. Associated with this atomic movement there is an
electron dynamics that can be interpreted as a successive closing
and opening of the step-edge gap or, alternatively, as an
increasing width of the step-edge energy levels. Therefore we can
expect changes in the spectral weight at E$_F$ as the temperature
is increased.

In summary, we report on first-principles DFT calculations
unveiling the origin of several
phenomena observed in the Si(557)-Au surface.
The experimental electronic band structure is successfully reproduced,
pointing out the important role played by the spin-orbit coupling in this
system. From a general perspective,
this raises the question of whether systems composed
by atomic-scale wires of heavy atoms on semiconducting
substrates can be used to create or
transport spin-polarized currents,
and thus be useful for future electronic devices.
The temperature dependence of the STM images
can be easily explained by the presence of two
structurally inequivalent silicon atoms in the step-edge
that fluctuate at room temperature between two positions.
The presence of a electronic gap
slightly above E$_F$ together with the step-edge fluctuation can
also explain
large changes in the
bands near the Fermi energy as a function of temperature.

\begin{acknowledgments}
The authors acknowledge useful discussions with P. Ordej\'on, J. E.
Ortega and
F. J. Himpsel. This work was supported by Basque Departamento
de Educaci\'on, the UPV/EHU (Grant. No.
9/UPV 00206.215-13639/2001), and the Spanish MCyT (Grant No.
MAT2001-09046).
DSP acknowledges support from the Spanish MCyT and CSIC.
SR acknowledges
support from the Max Planck Research Awards Fund, DIPC and 
the Emil Aaltonen foundation.
\end{acknowledgments}


\begin{references}
\bibitem{note} Note that in
Ref.~\onlinecite{SurfSci03} this assignment was erroneously reversed.
We are preparing an erratum pointing out this mistake.
\end{references}
\end{document}